\documentclass[12pt]{article}
\usepackage[latin1]{inputenc}
\usepackage{pstricks}
\usepackage{amsmath}
\usepackage{float}
\usepackage{color}
\input epsf
\usepackage{amssymb}
\usepackage[dvips]{graphicx,psfrag}
\newcommand{\be}{\begin{equation}}
\newcommand{\ee}{\end{equation}}
\newcommand{\bea}{\begin{eqnarray}}
\newcommand{\eea}{\end{eqnarray}}
\newcommand{\lb}{\label}
\begin{document}
\begin{titlepage}
\title{On the growth of linear perturbations}
\author{David Polarski\thanks{email:polarski@lpta.univ-montp2.fr}~
  and Radouane Gannouji\thanks{email:gannouji@lpta.univ-montp2.fr}\\
\hfill\\
Lab. de Physique Th\'eorique et Astroparticules, CNRS\\ 
Universit\'e Montpellier II, France}
\pagestyle{plain}
\date{\today}

\maketitle
\thispagestyle{empty}
\begin{abstract}
We consider the linear growth of matter perturbations in various dark 
energy (DE) models. We show the existence of a constraint valid at $z=0$ 
between the background and dark energy parameters and the 
matter perturbations growth parameters. 
For $\Lambda$CDM $\gamma'_0\equiv \frac{d\gamma}{dz}|_0$ lies in a very narrow interval 
$-0.0195 \le \gamma'_0 \le -0.0157$ for $0.2 \le \Omega_{m,0}\le 0.35$.
Models with a constant equation of state inside General Relativity (GR) are characterized by a 
quasi-constant $\gamma'_0$, for $\Omega_{m,0}=0.3$ for example we have 
$\gamma'_0\approx -0.02$ while $\gamma_0$ can have a nonnegligible variation. 
A smoothly varying equation of state inside GR does not produce either 
$|\gamma'_0|>0.02$. 
A measurement of $\gamma(z)$ on small redshifts could 
help discriminate between various DE models even if their $\gamma_0$ is close, 
a possibility interesting for DE models outside GR for which a significant 
$\gamma'_0$ can be obtained. 
\end{abstract}

\end{titlepage}

\section{Introduction}
There is growing observational evidence for the late-time accelerated expansion of our 
universe \cite{SS00}. This radical departure from conventional decelerated expansion is 
certainly a major challenge to cosmology. This non standard expansion could be due to an 
exotic non clustered component yet to be determined with a sufficiently negative pressure 
called Dark Energy (DE). By analogy all models trying to explain the accelerated expansion 
are called DE models but many models go now well beyond this simple picture. 
While the usual Friedmann equations in the presence of a cosmological constant term $\Lambda$ 
seem to be in good agreement with the data, it is clear that other models with a variable 
equation of state are allowed as well \cite{SS00}. While a cosmological constant universe is 
appealing because of its simplicity it nonetheless poses the problem of the magnitude of the 
cosmological constant $\Lambda$. This is the basic incentive to look for other models where 
DE has a variable equation of state. An additional incentive comes from the possibility to have 
phantom dark energy at low redshifts as this excludes the quintessence models, models with a 
minimally coupled scalar field inside GR \cite{RPW88}. 
It might also be that one should change the theory of gravity, as for example in scalar-tensor 
models \cite{ST,BEPS00}, and a lot of research has focused recently on other modified gravity 
models and higher dimensional models. 
Sometimes, the background expansion going 
back to high redshifts is enough to rule out some models \cite{APT07}, but typically this is not 
the case: models of a very different kind will be able to have a viable background expansion where 
the low redfshift expansion is in accordance with SNIa data.   

Depending on the gravity theory one is considering, the growth of the perturbations, even at 
the linear level, will be affected. Indeed, while distance luminosity measurements probe 
the cosmic expansion, matter perturbations probe in a independent way (see e.g. \cite{B06}) 
the gravity theory 
responsible for their growth (and of course also for the cosmic expansion). The growth rate 
of matter perturbations could be probed with three dimensional weak lensing surveys 
(see e.g. \cite{HKV07}). 
Hence two DE models based on different gravitation theories can give the same late-time 
accelerated expansion and still differ in the matter perturbations they produce \cite{S98}. 
This fact could provide an additional important way to discriminate between various models 
(see e.g. \cite{HL06}) and it is therefore important to characterize as accurately as possible 
the growth of matter perturbations which is the aim of the present work.

\section{Linear growth of perturbations}
Let us consider the dynamics of the linear matter perturbations. 
These perturbations satisfy a modified equation of the type 
\be
{\ddot \delta_m} + 2H {\dot \delta_m} - 4\pi G_{\rm eff}~\rho_m~\delta_m = 0~,\label{del}
\ee
The gravitational constant $G_{\rm eff}$ depends on the specific model under consideration 
and the corresponding modification of gravity. For example, as was shown in \cite{BEPS00}, for 
scalar-tensor DE models we have 
\be
G_{\rm eff} = G_N ~\frac{F+2(dF/d\Phi)^2}{F+\frac{3}{2}(dF/d\Phi )^2}
                = G_N ~\frac{1 + 2 \omega^{-1}_{BD}}{1 + \frac{3}{2} \omega^{-1}_{BD}} ~.\lb{GeffST}
\ee
An equation similar to (\ref{del}) is also found for example in DGP models. The physics behind it is 
a modification of Poisson's equation (see e.g. \cite{EP01}) according to (we drop the subscript $m$)
\be
\frac{k^2}{a^2}~\phi = -4\pi~G~\rho~\delta  \to  
               \frac{k^2}{a^2}~\phi = -4\pi~G_{\rm eff}~\rho~\delta~.\lb{Poisson}
\ee 
Of course, more drastic modifications are possible as well. In particular more elaborate DE models 
can be considered that could further increase the degeneracy between models inside and outside GR 
(see e.g. \cite{KS07}).  
It is convenient to introduce the quantity $f=\frac{d \ln \delta}{d \ln a}$. 
Then the linear perturbations obey the equation 
\be
\frac{df}{dx} + f^2 + \frac{1}{2} \left(1 - \frac{d \ln \Omega_m}{dx} \right) f = 
                              \frac{3}{2} \frac{G_{\rm eff}}{G_{N,0}} \Omega_m~.\lb{df}
\ee
with $x\equiv \ln a$. Equation (\ref{df}) reduces to eq.(B7) given in \cite{WS98} for 
$\frac{G_{\rm eff}}{G_{N,0}}=1$. The quantity $\delta$ is easily recovered using $f$ 
as follows 
\be
\delta(a) = \delta_i~{\rm exp} \left[ \int_{x_i}^{x} f(x') dx' \right]~.
\ee
We see that $f=p$ when $\delta\propto a^p$, in particular $f\to 1$ in $\Lambda$CDM for large $z$ 
and $f=1$ in an Einstein-de Sitter universe. 

An important issue is to characterize departures on small redshifts for different models. 
It is well known that for in a $\Lambda$CDM universe one can write 
\be
f \simeq \Omega_m^{\gamma}~,\lb{Omgamma}
\ee
with $\gamma =~{\rm constant}\sim 0.6$, an approach pioneered some time ago \cite{P84} and 
generalised in \cite{LLPR91}. The characterization of the growth of matter perturbations 
using a parametrization of the form (\ref{Omgamma}) has attracted a lot of interest in the 
hope to discriminate between DE models based on different gravity theories.   

Of course it is possible to write in full generality
\be
f=\Omega_m(z)^{\gamma(z)}~.  
\ee
Let us consider the quantity $\gamma'\equiv \frac{d\gamma}{dz}$.
For many models it turns out that
\be
\gamma(z)\approx \gamma_0 + \gamma'_0~z~~~~~~~~~~~~~~~~~~~~~~~0\le z\le 0.5~.\lb{gammaz}
\ee 
As we will see later, this could have interesting observational consequences. 

We now derive a constraint which is valid in general for any $\gamma(z)$. 
It is easy to obtain the following equation
\be
-(1+z)~\ln \Omega_m ~\gamma' + \Omega^{\gamma}_m +  \frac{1}{2}(1 + 3(2\gamma -1)~w_{\rm eff}) = 
                           \frac{3}{2}  ~\frac{G_{\rm eff}(z)}{G_{N,0}}~\Omega^{1-\gamma}_m~,\lb{dgamma}
\ee 
where $w_{\rm eff}\equiv w_{DE}~\Omega_{DE}$. From (\ref{dgamma}), it is easy to derive the following 
equation 
\be
\gamma'_0 = \left[ \ln \Omega^{-1}_{m,0} \right]^{-1} ~\left[ -\Omega^{\gamma_0}_{m,0} - 
   3(\gamma_0 - \frac{1}{2})~w_{{\rm eff},0} + \frac{3}{2}~\frac{G_{{\rm eff},0}}{G_{N,0}} 
                                     \Omega^{1-\gamma_0}_{m,0} - \frac{1}{2}\right]~.
\lb{dgamma0}
\ee
Equation (\ref{dgamma0}) is further simplified in models for which $\frac{G_{{\rm eff},0}}{G_{N,0}}=1$ 
to very high accuracy. An example where this is the case is provided by scalar-tensor DE models for 
which $0<\frac{G_{{\rm eff},0}}{G_{N,0}}-1< 1.25\times 10^{-5}$.  
We then obtain
\be
\gamma'_0 = \left[ \ln \Omega^{-1}_{m,0} \right]^{-1} ~\left[ -\Omega^{\gamma_0}_{m,0} - 
   3(\gamma_0 - \frac{1}{2})~w_{{\rm eff},0} + \frac{3}{2}~\Omega^{1-\gamma_0}_{m,0} - \frac{1}{2}\right]~.
\lb{dgamma0b}
\ee
This does not mean that equation (\ref{dgamma0b}) cannot differentiate between different 
gravitation theories satisfying $\frac{G_{{\rm eff},0}}{G_{N,0}}=1$ but rather that if it 
does so it is through the value of $\gamma_0$. This value is of course affected by the function 
$G_{\rm eff}(z)$. We will assume below $\frac{G_{{\rm eff},0}}{G_{N,0}}=1$ to very high accuracy. 
As we see from (\ref{dgamma0b}), we have $\gamma'_0 = \gamma'_0(\gamma_0,~\Omega_{m,0},~w_{DE,0})$ 
which is clearly equivalent to a constraint of the form
\be
f(\gamma_0,~\gamma'_0,~\Omega_{m,0},~w_{DE,0})=0~.\lb{f0}
\ee
In this connection one should note that fitting functions of $\gamma(z)$ proposed 
in the literature, even though they give a satisfactory fit for $f(z)$ in models 
satisfying some assumptions, generically will not satisfy the constraint (\ref{f0}).  
In contrast the constraint (\ref{f0}) does not depend on any assumption about $w(z)$.  
For fixed $\Omega_{m,0},~w_{DE,0}$, there will be a value $\gamma_{0,cr}$ for which 
$\gamma'_0=0$. However we will have generically $\gamma_0\ne \gamma_{0,cr}$ and 
therefore $\gamma'_0\ne 0$.  

Very generally, in any model for which the parameters $\Omega_{m,0}$ and $w_{DE}$ (and 
hence $w_{DE,0}$) are given, one can compute numerically the function $\gamma(z)$ from 
the linear growth of the matter perturbations. Using (\ref{f0}) it is then possible to 
obtain $\gamma'_0$. We will do this in the next Section for various models inside GR. 

Before considering specific DE models, it is possible to derive some general 
consequences from the constraint (\ref{f0}). Generically $\gamma'_0$ will not 
vanish, it needs not even be small. 
Let us consider $\gamma'_0$ in function of $\gamma_0$ for $\Omega_{m,0}$ and 
$w_{DE,0}$ fixed. 
As we can see from Figure 1a, the constraint (\ref{f0}) implies in excellent 
approximation a linear relation as follows 
\be
\gamma'_0 \simeq c + b~(\gamma_0 - 0.5)~~~~~~~~~~~~~~~~~~~~~~~
                                                      ~b\sim 3~. \lb{dgamma0f}
\ee 
The coefficients $c,~b$ depend on the background parameters $b=b(w_{DE,0},~\Omega_{m,0})$ 
(remembering that we take $\frac{G_{{\rm eff},0}}{G_{N,0}}=1$). 
The coefficient $b$ decreases while $c$ increases when $\Omega_{m,0}$ decreases from $0.35$ 
to $0.20$ (see Figure 1b). In contrast, $c$ increases from $-0.19$ for $\Omega_{m,0}=0.3$, 
to $-0.17$ for $\Omega_{m,0}=0.2$

For $\Omega_{m,0}=0.3$ we have $c=-0.19$. 
We stress that relation (\ref{dgamma0f}) will hold independently 
of any particular model and is a consequence of the constraint (\ref{f0}).

Depending on the specific model under consideration, for given background parameters 
$\Omega_{m,0}$ and $w_{DE,0}$, $\gamma'_0$ will take the value $\gamma'_0(\gamma_0)$ 
corresponding to the value $\gamma_0$ ``realized'' by the model. Generically we will 
have $\gamma'_0\ne 0$.
\begin{figure}[H]
\begin{centering}\includegraphics[scale=1.]{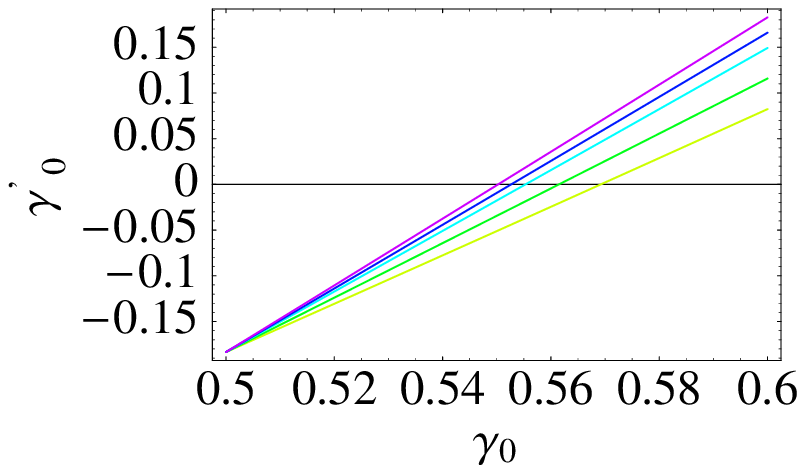}~~
\includegraphics[scale=.8]{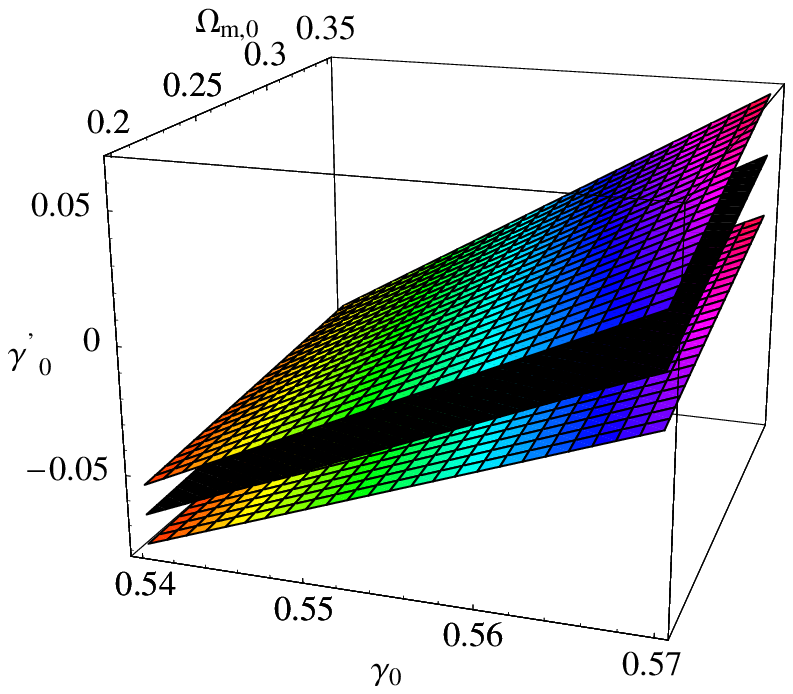}
\par\end{centering}
\caption{a) The left panel shows the constraint (\ref{dgamma0b}) for 
$\Omega_{m,0}=0.3$ and various values of $w_{DE,0}$. We have from top to bottom: 
$w_{DE,0}=-1.4,~-1.3,~-1.2,~-1,~-0.8$. For given  $\Omega_{m,0}$ and $w_{DE,0}$, 
the couple $\gamma_0,~\gamma'_0$ is on the corresponding line for any model while 
$\gamma'_0$ will depend on the value $\gamma_0$ realized in a particular model. 
b) On the right panel the constraint (\ref{dgamma0b}) is shown in function of 
$\Omega_{m,0}$. From top to bottom we have $w_{DE,0}=-1.2,~-1,~-0.8$. We see that 
the coefficient $b$ defined in (\ref{dgamma0f}) increases for increasing 
$\Omega_{m,0}$ and decreasing $w_{DE,0}$.}
\lb{Fig1}
\end{figure}
It is also seen from Figure 1a that a small variation of $\gamma_0$, for \emph{fixed} 
parameters $w_{DE,0},~\Omega_{m,0}$, can induce a non negligible variation of $\gamma'_0$ 
in accordance with eq.(\ref{dgamma0f}). In particular the relative change in $\gamma'_0$ 
can be very large. We will show below that for $w_{DE}=$ constant, the $\gamma'_0$ 
values are restricted to a very narrow range with $\gamma'_0\approx -0.02$. Even when one 
consider a smoothly varying equation of state, we still have 
$-0.02\lesssim \gamma'_0\lesssim 0.005$ (see below) for $0.20\le \Omega_{m,0}\le 0.35$. 
In other words a smooth change in the equation  of state of DE is not able to produce 
$\gamma'_0<-0.02$ for viable cosmological parameters.    
Therefore, a measurement of $\gamma'_0$ outside this range could be a characteristic 
signature of a DE model where gravity is modified. 
Moreover, a precise determination of $\gamma'_0$ could help to better discriminate between 
various modified gravity models.    

When $\Omega_{m,0}=0.3$ we have $b=3.13$ for $w_{DE,0}=-1$, while $b$ becomes smaller 
for $w_{DE,0}>-1$ and larger for $w_{DE,0}<-1$ (phantom DE today). 
Hence for $w_{DE,0}<-1$, we get a larger 
variation $\Delta \gamma'_0$ for a given variation $\Delta \gamma_0$.
When $\Omega_{m,0}$ decreases, so does the coefficient $b$ however this decrease is rather 
small for relevant cosmological values. 
It would be most interesting to investigate whether a precise determination 
of $\gamma'_0$ is observationally accessible. In view of (\ref{gammaz}) this means that one 
should measure precisely $\gamma(z)$ on $0\le z\le 0.5$. 
Another aspect concerns the extraction of one, or both, of the parameters $\Omega_{m,0}$ or 
$w_{DE,0}$. If we assume erroneously that $\gamma'_0=0$, a large error can result in the 
determination of $\Omega_{m,0}$ or $w_{DE,0}$ from the knowledge of $\gamma_0$. This is 
illustrated in Figure 2a.

\begin{figure}[H]
\begin{centering}\includegraphics[scale=1.]{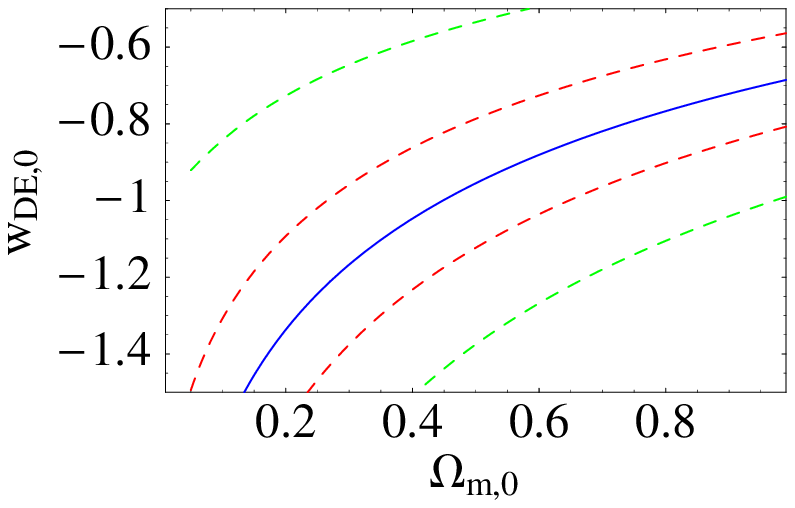}~~
\includegraphics[scale=1.]{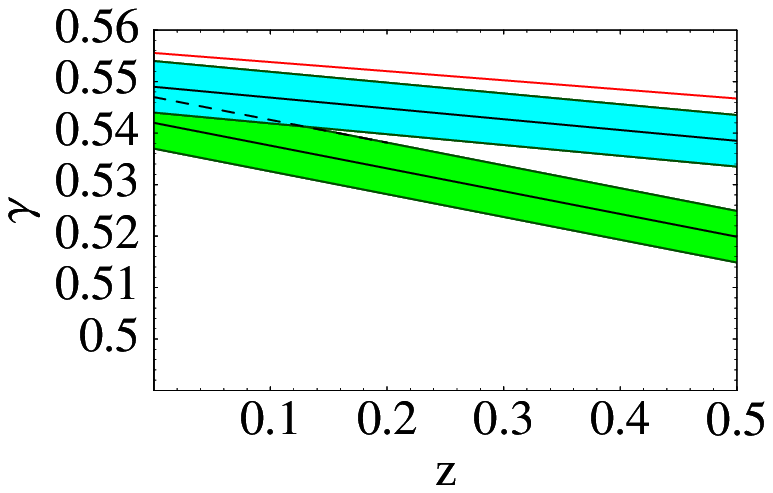}
\par\end{centering}
\caption{a) On the left panel, the blue line shows the degeneracy in the 
$\Omega_{m,0},~w_{DE,0}$ plane for $\gamma_0=0.555$ {\emph assuming} 
$\gamma'_0=0$. The red, resp. green, dashed lines correspond to $\gamma'_0=-0.02$ 
(top) and $\gamma'_0=0.02$ (bottom), resp. $\gamma'_0=-0.05$ (top) and 
$\gamma'_0=0.05$ (bottom). 
Ignoring the true non vanishing value of $\gamma'_0$ increases 
significantly the uncertainty on the couples $\Omega_{m,0},~w_{DE,0}$. 
b) On the right panel it is seen that models with very close $\gamma_0$ can be 
discriminated if $\gamma$ is measured for $0\le z\le 0.5$ assuming $\gamma$ is 
linear on small $z$, as often is the case. 
The lower the values of $\gamma_0$, the easier it is to discriminate these models 
through the difference in their slope $\gamma'_0$. 
For illustration, we have assumed here an error of $1\%$.} 
\lb{Fig2}
\end{figure}

\section{Some specific models}
We now turn our attention to specific models \emph{inside} General Relativity where DE has 
a known equation of state.   

\subsection{$\Lambda$CDM}
Because of its simplicity and of the recent data that seem to imply that viable DE models should 
not be too far from $\Lambda$CDM (see however \cite{Perc07}), this model plays a central role. 
We find for $\Lambda$CDM $0.554 \le \gamma_0 \le 0.558$ (see Figure 3b) and  
$-0.0195 \le \gamma'_0 \le -0.0157$ for $0.2 \le \Omega_{m,0}\le 0.35$. 
Hence $\gamma_0$ varies very little in function of $\Omega_{m,0}$ while $\gamma'_0$ 
is negative with $|\gamma'_0|<0.02$. 
An observation outside these values, in particular a positive value for $\gamma'_0$, 
or a large negative $\gamma'_0$, would signal a departure from $\Lambda$CDM.

\subsection{Constant equation of state}
We consider now a constant equation of state which includes of course the $\Lambda$CDM model. 
For the conservative ranges $0.2\le \Omega_{m,0}\le 0.35$ and $-1.5\le w_{DE,0}\le -0.5$, 
we find $0.542 < \gamma_0 < 0.583$ and $-0.021 < \gamma'_0 < -0.013$. 
However, as can be seen from Figure 3a, for fixed parameter $\Omega_{m,0}$, the value of $\gamma'_0$ 
is practically constant with $\gamma'_0\approx -0.02$ for different constant $w_{DE}$ despite a 
non negligible variation of $\gamma_0$. 
To summarize, for constant $w_{DE}$, $\gamma'_0$ lies in the restricted range $-0.024 <\gamma'_0<0.01$ 
while it is practrically constant if $\Omega_{m,0}$ is fixed. 
However, as emphasized above (see Figure 2a), even in that case neglecting the true 
(nonzero) value of $\gamma'_0$ can induce a significant error in the determination of $\Omega_{m,0}$ 
or $w_{DE,0}$ from $\gamma_0$. Finally, it is interesting to note that for given $\Omega_{m,0}$ 
all these models have essentially the same $\gamma'_0$ while the parameter $\gamma_0$ can vary 
by about $4\%$ (see Figure 3a).

\begin{figure}[H]
\begin{centering}\includegraphics[scale=1.]{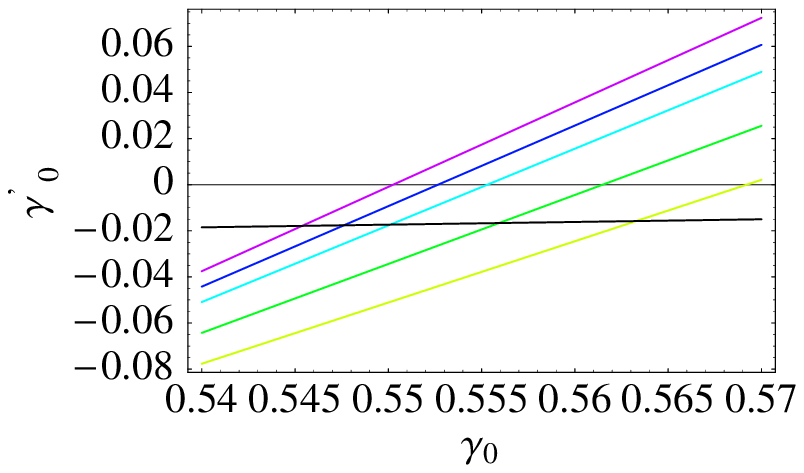}~~
\includegraphics[scale=1.]{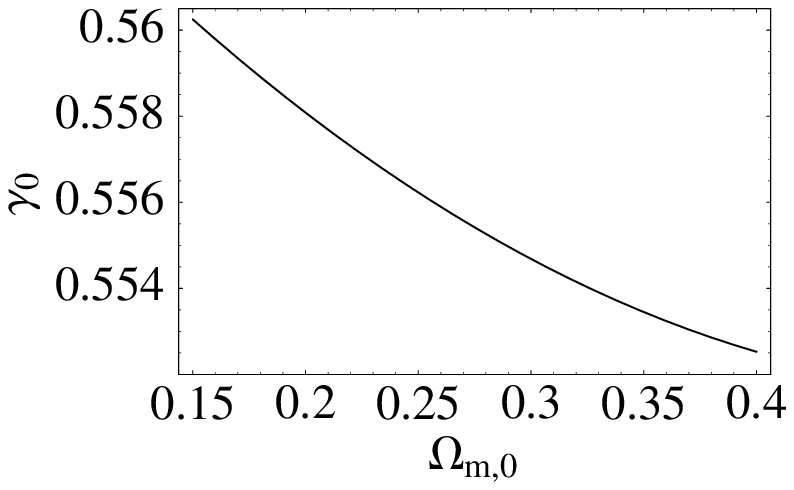}
\par\end{centering}
\caption{a) The lines in colour on the left panel are the same as in Figure 1. The black line 
gives the true value of $\gamma_0$ realised in models with $w_{DE}=w_{DE,0}$ = constant and 
$\Omega_{m,0}=0.3$. It is seen that all models with $w_{DE}$ = constant shown here have 
practically the same non vanishing $\gamma'_0$, $\gamma'_0\approx -0.02$. Note that 
$\gamma_0$ increases when $w_{DE}$ increases.
b) On the right, $\gamma_0$ is displayed in function of $\Omega_{m,0}$ for the $\Lambda$CDM model.}
\lb{Fig3}
\end{figure}

\subsection{Variable equation of state}
Our analysis can be repeated for DE with a variable equation of state. To be 
specific, we take a smoothly varying equation of state of the type \cite{CP01,L03}
\be
w_{DE}(z)  =  (-1 + \alpha) + \beta ~(1-x) \equiv w_0 + w_1~\frac{z}{1+z}~,\\ \lb{wCPL}
\ee
where $x\equiv \frac{a}{a_0}$. The corresponding evolution of the DE energy density can be 
computed analytically and yields \cite{CP01}
\be
\rho_{DE}(z) = \rho_{DE,0}~(1+z)^{3(\alpha + \beta)} {\rm e}^{-3\beta \frac{z}{1+z}}~.\lb{epsCPL}
\ee
The results are displayed in Figure 4 for models with a negligible 
$\Omega_{DE}$ for $z\gg 1$. For example, if we fix $w_0=-1.2$, we can 
compute the values of $\gamma_0$ and $\gamma'_0$ in function of 
$\beta\equiv w_1$ and $\Omega_{m,0}$. We find 
$0.55\lesssim \gamma_0\lesssim 0.56$ and $-0.022\lesssim \gamma'_0\lesssim 0.005$ 
for $0.20\le \Omega_{m,0}\le 0.35$ and $0\le \beta\le 1$. 
\footnote{Note that slightly lower values for $\gamma'_0$ can be obtained for less 
interesting models with substantial phantomness in the asymptotic past}. 
To summarize, a smoothly varying equation of state does not seem able to 
generate $|\gamma'_0|> 0.02$. 

\begin{figure}[H]
\begin{centering}\includegraphics[scale=.8]{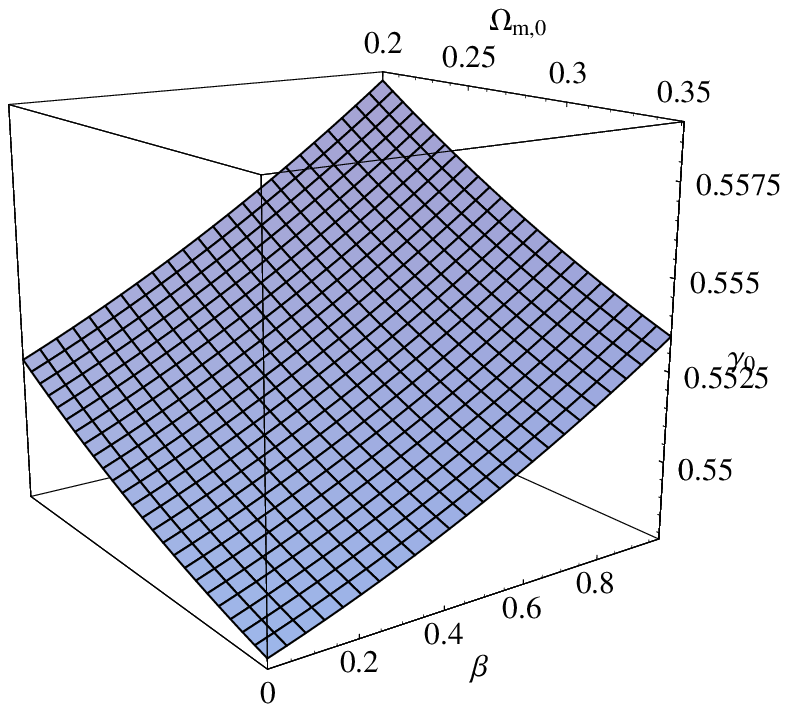}~~
\includegraphics[scale=.8]{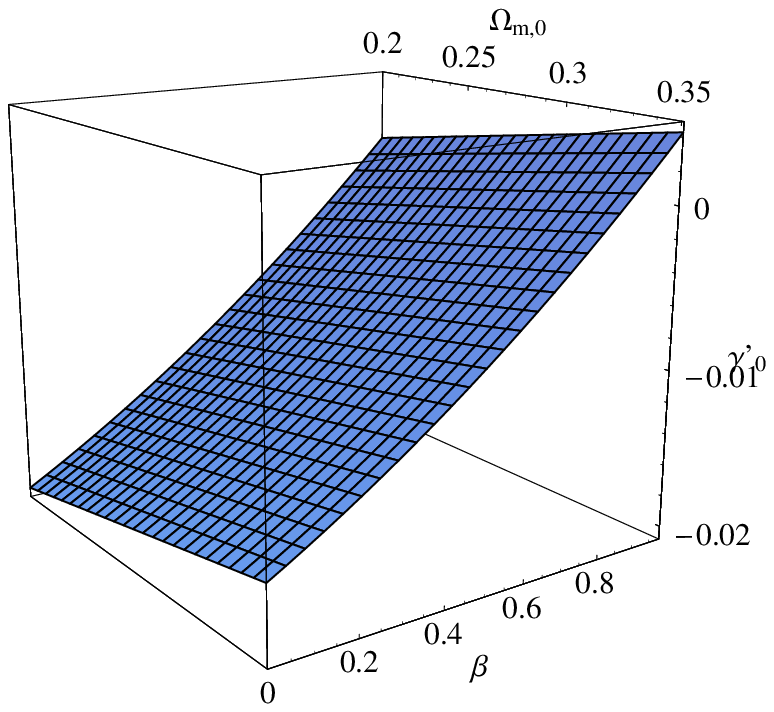}
\par\end{centering}
\caption{The parameters $\gamma_0$ (left) and $\gamma'_0$ (right) are shown 
in function of $\beta\equiv w_1$ and $\Omega_{m,0}$ for a model with variable 
equation of state parameter $w_{DE}= -1.2 + \beta~\frac{z}{1+z}$. Hence all 
the points on the two surfaces have $w_{DE,0}=-1.2$. The results for 
$w_{DE}=-1.2$ are recovered for $\beta=0$. We note for the left figure that 
$\gamma'_0=0$ is obtained for some particular combinations $\beta,~\Omega_{m,0}$.}
\lb{Fig4}
\end{figure}

\section{Summary and conclusions}
Considering the linear growth of matter perturbations in various models, we give a constraint 
at $z=0$, eq.(\ref{dgamma0b}) valid for all models, including modified gravity DE models that 
satisfy $\frac{G_{{\rm eff},0}}{G_{N,0}}=1$. 
This constraint implies that the quantity $\gamma'_0$ is completely fixed by the 
remaining parameters $\gamma_0,~w_{DE,0}$ and $\Omega_{m,0}$. 
For the models considered here inside GR, $|\gamma'_0|\lesssim 0.02$. Interestingly for 
models inside GR with constant $w_{DE}$, $\gamma'_0$ is quasi-constant with 
$\gamma'_0\approx -0.02$ as the variation of $w_{DE,0}$ is compensated by a simultaneous 
variation of $\gamma_0$ (for given $\Omega_{m,0}$). 

We have generically 
$\gamma'_0\ne 0$ and we emphasize that a significant $\gamma'_0$ could help discriminate 
between models, even if their $\gamma_0$ values are close. We have illustrated this 
schematically on Figure 2b.
This potential resolution improves as $\Omega_{m,0}$ goes up and/or $w_{DE,0}$ goes down 
and could be important when dealing with DE models outside General Relativity. 
We will give elsewhere specific models where this is the case \cite{GP07}.
Generally, this approach could be very fruitful whenever $\gamma(z)$ is close to linear 
on small redshifts $0\le z\le 0.5$ so that the slope is essentially given by $\gamma'_0$. 
So we feel it would be useful to try to measure $\gamma(z)$ on small redshifts, and not 
just $\gamma_0$. 
Finally it is important to realize that neglecting a small but nonvanishing $\gamma'_0$ can 
induce a large error on the parameters $\Omega_{m,0},~w_{DE,0}$ that one could extract from 
the growth of matter perturbations. 


\end{document}